# Using agent-based models and EXplainable Artificial Intelligence (XAI) to simulate social behaviors and policy intervention scenarios: A case study of private well users in Ireland

Rabia Asghar[1,3], Simon Mooney[2], Eoin O'Neill[2], Paul Hynds[1,3]

[1]*Sustainability and Health Research Hub, Technological University (TU), Dublin, Ireland*
[2]*School of Architecture, Planning and Environmental Policy, University College Dublin (UCD), Dublin, Ireland*
[3]*Spatiotemporal Environmental Epidemiology Research (STEER) Group, Technological University (TU), Dublin, Ireland*
rabia.asghar@tudublin.ie, simon.mooney1@ucd.ie, paul.hynds@tudublin.ie, eoin.oneill@ucd.ie

***Abstract***—

Approximately 50% of the Republic of Ireland's (ROI) rural population relies on unregulated private wells which are susceptible to agricultural runoff and untreated domestic wastewater. Elevated national rates of Shiga toxin-producing *Escherichia coli* (STEC) and other waterborne illnesses have been increasingly linked with well water exposure. Therefore, promoting positive behavioural actions, such as periodic water testing, is vital for safeguarding rural public health. However, an absence of centrally incentivised water quality testing requires household expenditure. Understanding the environmental, cognitive, and material factors influencing these behaviours is essential.

Existing studies have paid little attention to the impacts of conjectural policy-based changes and inter-agent interactions, such as those between private well users and government agencies. To address this gap, the authors used an agent-based modelling (ABM) approach, representing a behavioural and legislative "sandbox" to simulate multiple future scenarios and subsequent behaviours based on national survey data. The ABM framework, parameterized to model private groundwater well-testing behaviours, incorporated a Deep Q-network reinforcement learning model and Explainable Artificial Intelligence (XAI) for interpretable insights on decision-making. Key features (model inputs), including weather, self-efficacy, and penalization/reward structures, were identified using Recursive Feature Elimination (RFE) with 10-fold cross-validation, while SHAP (Shapley Additive Explanations) values, a core XAI approach, further explained feature importance, enhancing ABM interpretability and facilitating actionable policy insights.

Overall, 14 hypothetical scenarios were developed and evaluated. The most effective intervention, "Free Well Testing + Communication Campaign," resulted in 435/561 agents participating in testing, from a current baseline of approximately 5%, achieving faster learning efficiency. "Free Well Testing + Regulation" also performed well, with 433/561 agents predicted to initiate well testing, demonstrating a similarly efficient adaptation rate. Free testing alone increased testing frequency from 5% to over 75%, with many residents testing multiple times annually, reinforcing the impact of well-testing incentives on behavioral change. The free well-testing scenarios consistently exhibited faster learning efficiency, converging within 1000 episodes, indicating that agents adapted more quickly to these interventions.

In contrast, other scenarios required approximately 2000 episodes, taking twice the time for decision-making, indicating slower adaptation and prolonged convergence periods. This study highlights the novel application of ABM and XAI in public health decision-making, providing a transferable framework for assessing interventions in various environmental and behavioral health contexts.

## 1. INTRODUCTION

Risk prevention behaviours at the household scale can have both immediate and lasting implications for the integrity of local aquatic environments [1–3]. Groundwater represents an aquatic environment subject to significant human influence, representing the principal source of private domestic drinking water in many rural regions globally [4, 5]. Appropriate implementation of private well management actions provide protection against ingress of microbial and chemical contaminants from anthropogenic activities, e.g., mismanagement of domestic wastewater treatment systems and misapplication of industrial substances [5, 6]. However, these actions require prior awareness and incur both time and financial costs, including acquisition and understanding of guidance information, assessment of supply vulnerability, and financial expenditure on (a) perennial treatment systems and (b) periodic water quality testing [7, 8]. Moreover, spatiotemporal triggers of supply contamination, such as extreme weather events (EWEs) and seasonal agro-industrial land use, require ongoing vigilance and continuity of household supply maintenance, highlighting the importance of strategic, rigorous well-testing promotion [9]. Previous behavioural research has consistently highlighted financial concerns, integrative capacity and cognitive factors (particularly awareness, risk perception and perceived control) as key drivers of both drinking water and generalised environmental protective actions [10–12]. As such, it is broadly understood that increased employment of multi-modal, facilitative interventions (e.g. combined public information campaigns and incentivised well-testing programmes) can substantially elevate rates of private well maintenance. While there is some evidence that risk communication campaigns (where appropriately tailored) and convenience-based interventions may positively motivate behaviour adoption [13, 14], the comparative efficacy of these initiatives has yet to be fully established.

The Republic of Ireland (ROI) represents a relevant case study for research in private groundwater management, considering its high levels of groundwater reliance [15]. Approximately 16% of the ROI's population (800,000 people or almost half of the ROI's rural population) uses a private well, a significantly larger proportion than other high-income European countries such as England and France. Moreover, the ROI's similarly high reliance on domestic wastewater treatment systems (DWWTSs), predominantly agricultural land-base and locally dense rural settlement patterns expose domestic wells to myriad contamination sources and pathways. As such, the potential gains conferred by improved communication and incentivisation of private well management towards the health of rural well owners may be significant – particularly considering an estimated 80% of annual cases of Shiga-toxin producing *Escherichia coli* (STEC) infection in the ROI are associated with well water consumption [16].

Given the absence of free supply testing, relative paucity of purposive communicative interventions and high rates of microbial groundwater contamination associated with the ROI, the likely impacts of future risk communication and policy interventions warrant consideration. Most modeling studies of key well-maintenance actions (e.g., testing) to date have notably focused on predicting levels of behavior adoption using "static" parameters and approaches. Little if any attention has been given to the potential impact of (a) scenario-based shifts on future behavior rates, (b) interactions between dynamic agents (i.e., well owners/users, local authorities), and (c) interactions between agents and dynamic environments. Agent-based modeling (AMB) represents a computer simulation approach that conceptualizes the actions, interactions, and responses of individual agents and their environment, thus providing a relevant modeling tool for behavior characterization in future hypothetical scenarios. This approach may enable the quantification of future behavior rates and the path-dependence of interventions. While ABM has begun to see usage in groundwater studies in a macro-management context [20], hypothetical simulations of groundwater management and environmental/cognitive stimuli at the household (private) level have yet to be explored. Accordingly, the authors have adopted ABMs to simulate the behavior of private groundwater well owners in the ROI using an existing survey-based dataset [12].

**This research's key contributions are:**

The explainable Agent-Based Model (ABM) framework was developed with two objectives:

1. To demonstrate how input changes impact private groundwater well-testing behaviors.
2. To encourage agents (i.e., well users) to test their groundwater wells more frequently.
3. To explore how Irish well users respond to different policy measures and identify effective ways to encourage more frequent well testing.

This research applies Reinforcement Learning (RL) to promote sustainable groundwater management by optimizing the decision-making process for 561 agents in the Irish dataset across two distinct Agent-Based Models (ABMs) focused on private groundwater well-testing.

The paper is organized as follows: Section II: Survey Design outlines the design and data collection process of the survey conducted. Section III: Proposed Methodology describes the agent-based modeling framework and reinforcement learning techniques employed in the study. Section IV: Performance Metrics explains the metrics used to evaluate the interventions and model performance. Section V: Results and Discussion presents the findings of the study and provides a detailed discussion of their implications. Finally, Section VI: Conclusion summarizes the key insights, policy implications, and recommendations for future research.

## 2. Survey Design

The survey followed a structured, standardized format informed by the KAP (knowledge, attitudes, and practice) model [21]. While the KAP model was not operationalized for the specific purpose of theory validation/hypothesis confirmation, the provision of questions broaching self-efficacy, attitudes, risk perception, and concern examined nevertheless allowed opportunities for theory exploration. The survey comprised 41 questions and queried private well users about previous experiences, well-protective actions, and scored cognitive features (i.e., risk perception and concern). The survey additionally broached the household history of gastrointestinal illness, supply awareness, and testing frequency as well as supply- and rural-specific variables such as household tenure with supply (i.e. residence during installation), supply type (i.e. dug or drilled), well water use (i.e. domestic/agricultural/irrigation-based) mode of domestic wastewater disposal (i.e., domestic wastewater treatment system or public sewer). Scoring frameworks used to formulate discrete variables (i.e., awareness, risk perception, groundwater beliefs, and attitudes) are outlined in [12]. By way of example, the awareness scoring framework is outlined in Table I.

**Table 1.** Awareness scoring frameworks used in Ontario and ROI surveys.

| Awareness domain | Response categories | | Scoring protocol | Assigned score |
|---|---|---|---|---|
| **Well age** | 0-5 years | 30-50 years | Aware | 1 |
| | 5-10 years | > 50 years | Unaware | 0 |
| | 10-20 years | Don't know | | |
| | 20-30 years | | | |
| **Well depth** | < 10 ft (3m)<br>10-50 ft (3-15m)<br>50-100 ft (15-30m)<br>100-200 ft (30-60m) | 200-300 ft (60-90m)<br>> 300 ft (90m)<br>Don't know | Aware<br>Unaware | 1<br>0 |
| **Well features * ǂ** | Well cap present<br>Damaged well cap<br>Pump at base of well | Cement well casing<br>Damaged well casing<br>Buried well | Aware<br>Unaware | 1<br>0 |
| **Treatment use** | Yes<br>No | Don't know | Aware<br>Unaware | 1<br>0 |
| **Previous test** | Yes<br>No | Don't know | Aware<br>Unaware | 1<br>0 |
| **Relevant pathogens** | *STEC*<br>*Giardia*<br>*Salmonella* | *Cryptosporidium*<br>*Campylobacter*<br>*Norovirus* | Aware of 5-6<br>Aware of 3-4<br>Aware of 1-2<br>Aware of none | 3<br>2<br>1<br>0 |
| **Pathogen sources * †** | Domestic animals<br>Grazing animals | Farmyards<br>Septic tanks | Aware of 3-4<br>Aware of 1-2<br>Aware of 0 | 2<br>1<br>0 |

[1] Maximum awareness score = 14

[2] Maximum awareness score = 12

### A. *Survey distribution:*

The survey was circulated both online (hosted via SurveyMonkey™) and in-person and was available from mid-September to late-November 2019. Survey dissemination was largely implemented online, with the electronic mode of survey completion prioritised due to superior cost-efficiency and response rates relative to physical surveys [22], [23]. The survey was advertised/distributed across multiple stakeholder groups and geographical units of administration to ensure maximum socio-demographic representativity. Surveys were circulated via a series of rural interest groups (e.g. community organisations, professional organisations), media platforms, government bodies and educational institutions after disclosure of study objectives, parameters and data management protection procedures.

## 3. Developed Methodology

The methodology integrates Agent-Based Modeling (ABM) with Deep Q-Networks (DQN) and Shapley Additive Explanations (SHAP) to enhance the explainability of Artificial Intelligence (XAI). Reinforcement Learning (RL) optimizes decision-making for 561 Irish agents in two ABMs focused on private groundwater well testing. In the first ABM, agents decide whether to test (0 or 1) based on environmental and socio-demographic attributes. Recursive Feature Elimination (RFE) and SHAP values identify the most influential features to ensure realistic decision-making. In the second ABM, agents who choose to test (action = 1) determine their testing frequency (ranging from 1 to 4). These trained baseline ABMs, using the top 20 feature sets, were then evaluated across 14 hypothetical scenarios, as illustrated in Figure 1.

This methodology section outlines data preprocessing, feature engineering, model interpretation (SHAP), explainable agent-based modeling (XABM), Agent Interactions, reinforcement learning, DQN model, and scenarios weighting. This framework contributed to developing a parameterized Agent-Based Model (ABM) and evaluating 14 hypothetical scenarios to optimize frequently groundwater well-testing decisions.

### *3.1. Data Preprocessing*

Features with incomplete (>30%) or interrupted data were excluded to maintain the dataset's integrity. Missing numerical values, with up to 10% missingness, were imputed using the median, chosen for its robustness to outliers, ensuring all features were fully populated. Categorical variables, such as demographics and postal codes, were transformed via one-hot encoding into binary vectors suitable for machine learning models. Outliers in continuous variables were detected using the Interquartile Range (IQR) method, identifying values outside the range of Q1 - 1.5 * IQR to Q3 + 1.5 * IQR. These outliers were flagged rather than removed to retain data while allowing further exploration. Continuous features were scaled using Standard Scaler, standardizing them to have a mean of 0 and a standard deviation of 1, ensuring all features contributed equally to model performance.

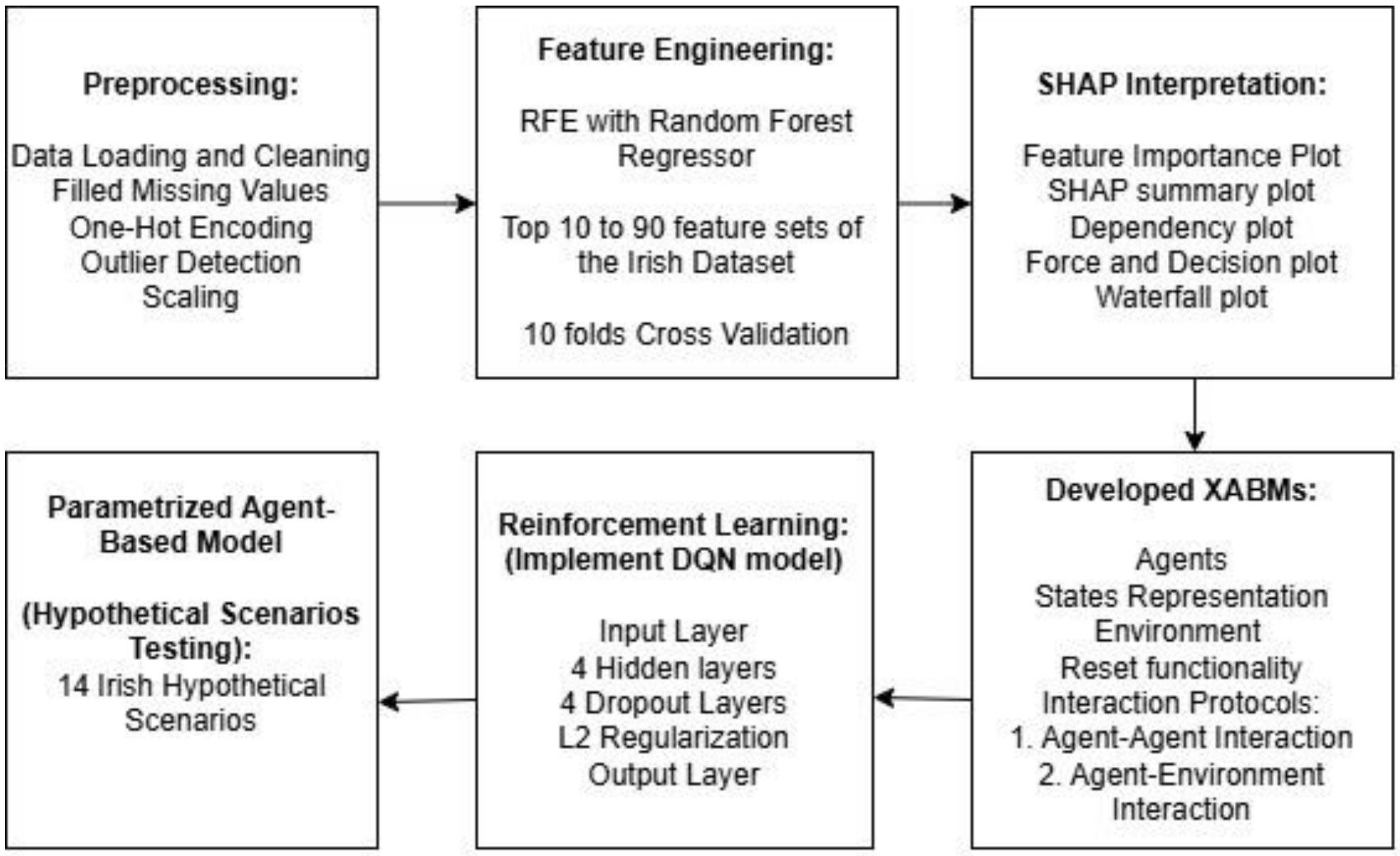

**Figure 1**. The methodological framework illustrates the following key steps: (1) Data preprocessing, which includes cleaning, encoding, and scaling; (2) Feature engineering using Recursive Feature Elimination (RFE) with a Random Forest Regressor for top feature selection; (3) SHAP-based feature interpretation, involving importance and dependency analysis; (4) Development of explainable agent-based models (XABMs), focusing on agent functionality, states, and interactions; (5) Reinforcement learning model implementation using a deep Q-network (DQN) structure; and (6) Parameterized agent-based modeling, involving the evaluation of 14 hypothetical scenarios

### *3.2. Feature Engineering*

A Random Forest Regressor with 100 estimators assessed feature importance, while Recursive Feature Elimination (RFE) iteratively removed less significant features. Feature subsets, ranging from the top 10 to the top 90, were tested to identify the optimal set. The selected features were then used to retrain models, improving decision-making performance and generalizability. The robustness of this approach was validated using 10-fold cross-validation, with 80% of the data for training and 20% for testing.

### *3.3. Shapley Additive exPlanations Interpretation*

Shapley Additive Explanations (SHAP) were used to interpret feature contributions in the Random Forest Regressor, enhancing model explainability within an Agent-Based Model (ABM). After applying Recursive Feature Elimination (RFE) with 10-fold cross-validation, SHAP values were computed for feature subsets ranging from 10 to 90 attributes to assess their individual impact. SHAP-based machine learning models were then applied to the Irish dataset, evaluating how each feature influenced model predictions—whether positively or negatively. This analysis identified the most significant feature set, providing valuable insights for policymaking and improving transparency in decision-making. The Python SHAP library generated key visualizations, including SHAP Summary and Dot Plots for feature importance.

### 3.4. Explainable Agent-Based Models (XABM)

Agent-based models (ABMs) simulate interactions among autonomous agents within a dynamic environment. This study represents 561 Irish agents based on survey responses from 561 Irish well users, integrating machine learning (ML) and deep learning (DL) to enhance their adaptive learning behaviors. The Explainable Agent-Based Model (XABM) integrates Explainable Artificial Intelligence (XAI) to improve transparency in AI-driven decision-making.

Each agent, defined by 10–90 attributes, independently makes well-testing decisions. The model systematically adjusts these attributes in increments to assess their influence on well-testing behavior. These 90 attributes determine state representation, directly shaping private well water testing decisions.

A reward-based system reinforces adaptive decision-making: positive rewards encourage timely testing during high-risk periods, such as autumn, while negative rewards/Penalty penalize missed or delayed tests, reflecting potential health and environmental consequences. The cumulative reward system incentivizes consistent and proactive well-testing behaviors, ensuring alignment with environmental and public health priorities.

Interaction dynamics occur through agent-agent engagements and agent-environment interactions, where seasonal factors influence agent actions. These interactions are:

#### 3.4.1. Agent-Agent Interactions:

This research developed a multi-agent system in which multiple agents operated within a shared environment. The actions of one agent not only influenced its own state transition but also impacted the state transitions of other agents. Specifically, when agent 1 took an action, it affected its own state transition $S_1^{n+1}$ and simultaneously influenced the state transition of another agent 2, i.e., $S_2^{n+1}$. This agent-to-agent interaction was implemented to accurately model interdependencies among 561 Irish agents, ensuring that their decisions dynamically adjusted in response to changes induced by other agents.

#### 3.4.2 Agent-to-Environment Interaction

Agent-to-environment interactions were modeled to capture the dynamic relationship between agents and environmental conditions. Each agent made testing decisions based on its current state $s^n$, which evolved over time due to external environmental factors such as seasons/months. The agent selected an action $a^n$, such as conducting a test at a specific well location, which resulted in a state transition:

$$S^n \rightarrow s^{n+1}$$

The environmental conditions influenced the risk of contamination and varied across different seasons. To adapt to these variations, agents dynamically adjusted their testing strategies as follows:

- **Winter (December–February)**: High rainfall (approximately 130 mm per month), but colder temperatures slow microbial activity and water movement, leading to lower testing frequency.
- **Spring (March-May)**: Moderate rainfall (around 100 mm per month) and rising temperatures enhance groundwater recharge, increasing the risk of pollutant leaching and necessitating moderate testing frequency.
- **Summer (June–August)**: This is the driest season, with rainfall averaging 80 mm monthly. Reduced contamination risk from surface runoff leads to the lowest testing frequency of the year.
- **Autumn (September–November)**: The highest testing frequency occurs due to heavy rainfall (approximately 130 mm per month) and intensified agricultural activities (e.g., soil tilling and fertilizer application), which increase the risk of groundwater contamination via surface runoff and leaching.

These seasonal adaptations ensured the agents efficiently monitored environmental conditions while optimizing testing efforts. By integrating real-time environmental data, the agents were able to adjust their decisions dynamically, improving the overall efficiency of contamination detection and resource management.

### 3.5. Reinforcement Learning and Deep Q-learning

To ensure optimal decision-making, agents utilized Reinforcement Learning (RL) with advanced Deep Q-learning and a shared replay memory (D) that stored both agent-to-agent and agent-to-environment interactions. This approach enabled agents to learn from past interactions by replaying stored data and extracting meaningful patterns. Through this feedback RL process, agents dynamically optimize their policies, improving well-testing decision-making in complex environments.

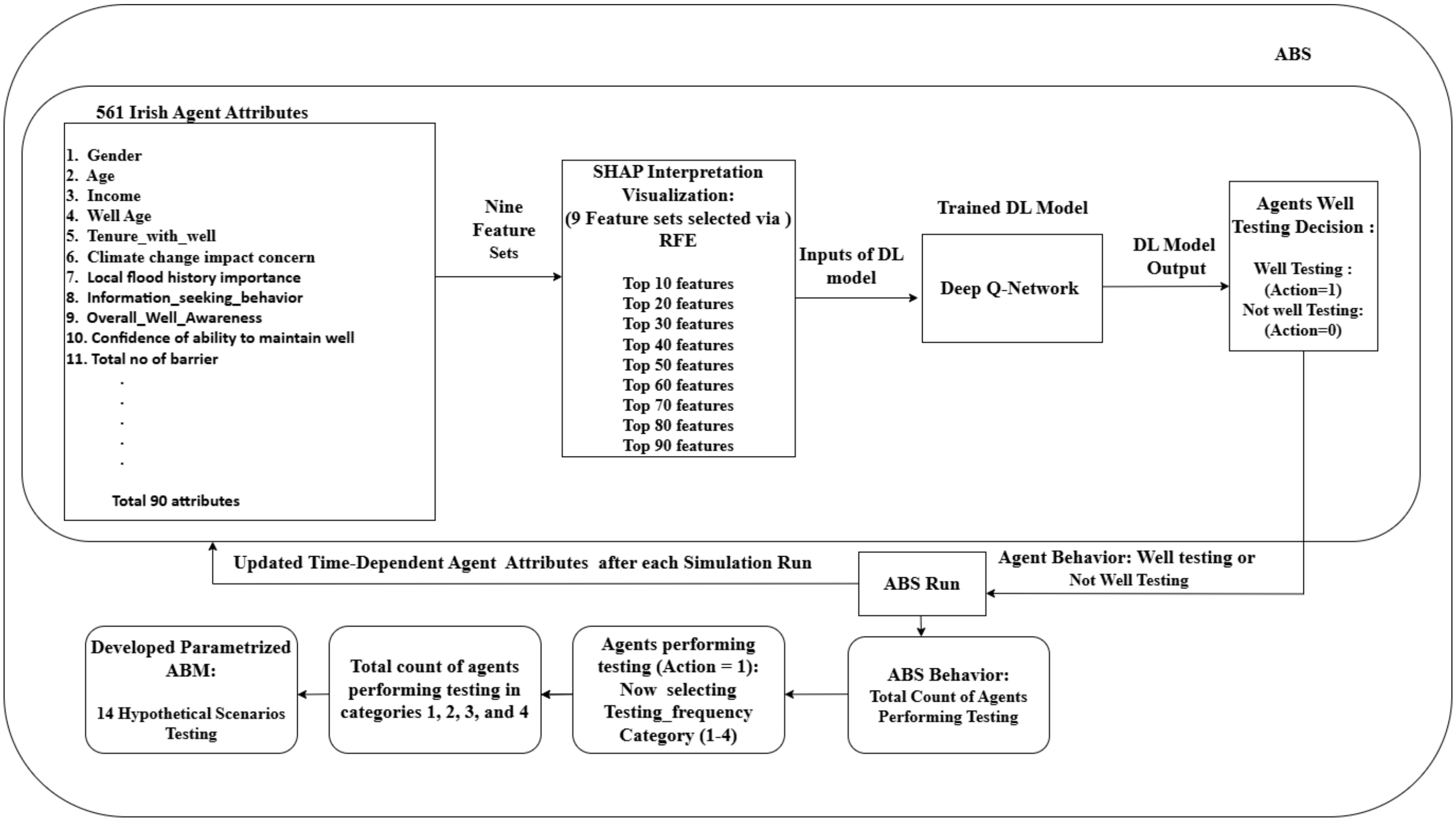

**Figure 2.** This workflow diagram illustrates the integration of agent-based modeling (ABM) with a Deep Q-network (DQN) for analyzing Irish agent behaviors in well-testing decision-making. The process involves selecting the top 9 feature sets via SHAP interpretation, training a DL model, and simulating agent decisions under 14 hypothetical scenarios. After each simulation run, dynamic updates to time-dependent agent attributes reflect the results.

### 3.6. Deep Q-Network (DQN) Architecture

The Deep Q-Network (DQN) model was implemented to optimize decision-making by learning the action-value function Q (s, a) through experience replay and target networks. The agent was modeled as a fully connected deep neural network (DNN) consisting of an input layer, four hidden layers, and an output layer. At each step, the network processes the observed state through the input layer, propagating the information through the hidden layers to compute Q-values for each possible action in the output layer. Given the current state, this network design allows the agent to estimate $Q*(s, a)$, the expected cumulative reward for each action. The workflow diagram, as shown in Figure 2, illustrates the integration of ABM with DQN to predict Irish agents' well-testing decision-making behaviors.

1) **Input Layer:** The input layer of the model processes the agent's state vector, which includes demographic data, behavioral patterns, and environmental attributes. The dimensionality of this layer corresponds to the total number of features in the state vector. To improve the model's understanding of the dataset, nine feature subsets were created for the Irish dataset, ranging from the top 10 to the top 90 features (state vectors), as determined by SHAP (XAI) analysis. This selection method ensured that the model prioritized the most influential features, optimizing its decision-making performance. The network functions as a deep neural network (DNN), where the input layer first encodes the observed state vector. The encoded vector is then passed sequentially through the hidden layers for further processing. The Deep Q-Network (DQN) architecture for groundwater well-testing decision-making is illustrated in Figure 3.
2) **Hidden Layers:** The architecture consists of four fully connected hidden layers designed to progressively enhance the model's ability to learn complex representations. These layers contain 64, 128, 256, and 512 nodes. During the forward pass, the input vector propagates through these layers, where each node computes a weighted sum of the inputs (dot product of weights and input vector) combined with a bias term. This value is then transformed using the ReLU activation function. The Rectified Linear Unit (ReLU) activation function is applied at each layer, introducing non-linearity and enabling the model to capture intricate patterns within the data. The output of each hidden layer serves as the input for the subsequent layer, progressively refining feature representations. To mitigate overfitting, dropout mechanisms are strategically applied, with dropout rates of 0.20, 0.25, and 0.15 across different layers.
3) **Output Layer:** The output layer generates Q-values for primary decision-making in the Agent-Based Model (ABM), with two neurons corresponding to the actions $\mathbf{a \in \{0,1\}}$. These neurons output Q-values $\mathbf{Q_n(s_n, 0)}$ and $\mathbf{Q_n(s_n, 1)}$, where $\mathbf{Q_n(s_n, 0)}$ represents the output node for those agents who are not performing a test (**a = 0**), and $\mathbf{Q_n(s_n, 1)}$ represents the second output node for agents who are performing a test (**a = 1**). Agents choosing a = 1 receive a positive reward (+1), reinforcing the testing behavior, while agents choosing a = 0 incur a penalty (-1) as a negative reward. The negative reward for **a = 0** is integrated into a feedback system through RL, enabling agents to learn adaptively, as shown in Figure 2.

   In the secondary ABM model, the output layer contains four actions representing different well-testing frequencies. For agents selecting **a = 1**, the specific testing frequency is denoted as **a_f ∈ {1, 2, 3, 4}**, corresponding to:
   - **a_f = 1** → Denotes individuals conducting well water testing 2–3 times yearly.
   - **a_f = 2** → Refers to individuals who test their well water **once annually**.
   - **a_f = 3** → Represents individuals who perform well testing **once every few years**.
   - **a_f = 4** → Identifies individuals who have **tested their well water only once**.

The reward hierarchy incentivizes agents to select well-testing frequencies based on a structured reward system. The highest reward (+2.5) is assigned to test frequency a_f = 3, followed by a_f = 4 with a reward of +2. a_f = 2 receives a reward of +1, and a_f = 1 is assigned a reward of +0.5. This framework simulates real-world constraints, where well-testing in Ireland is not provided for free. The model incorporates dynamic decision-making by agents, where their testing preferences are influenced by changes in the cost structure of testing, including scenarios with and without free testing.

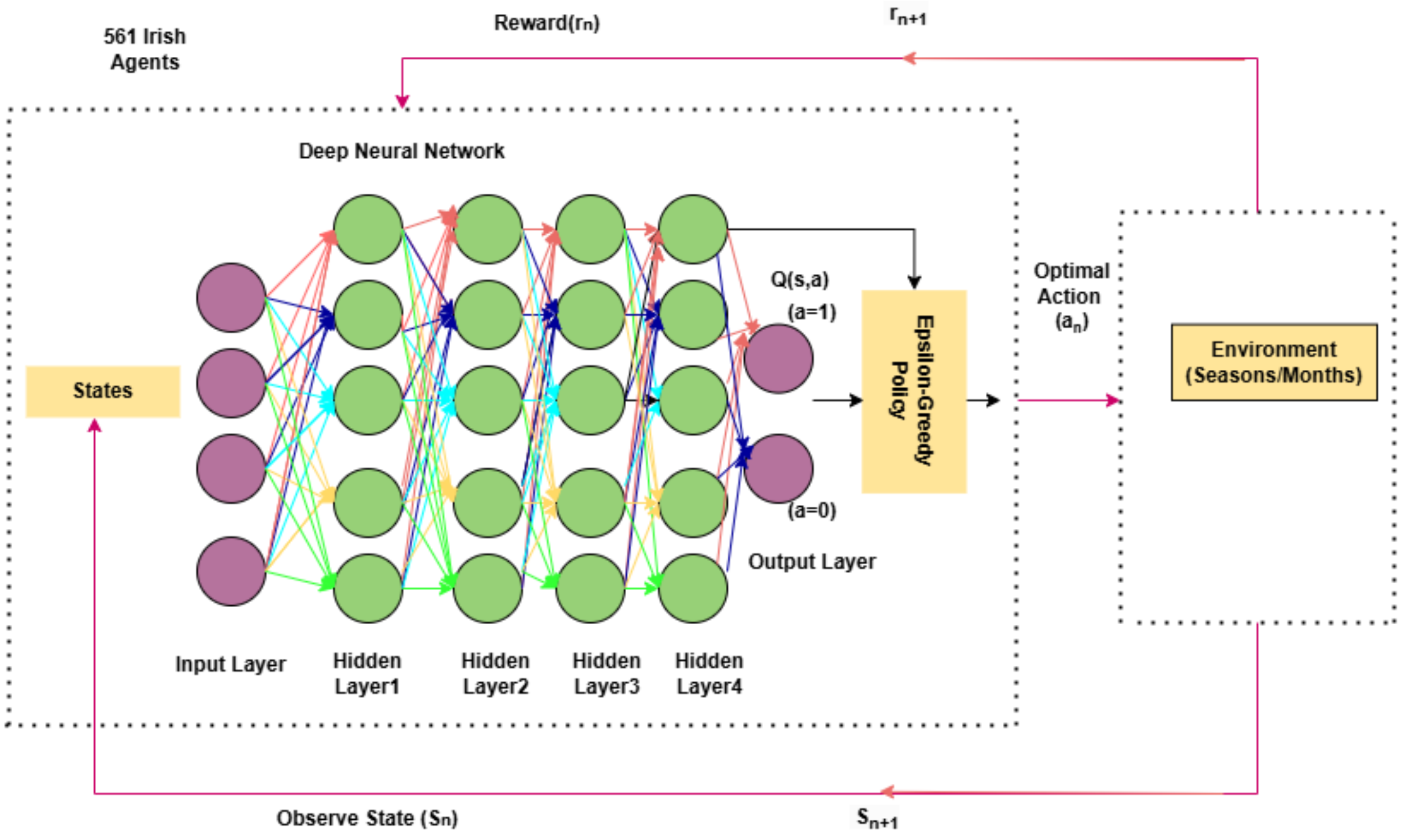

**Figure 3.** Deep Q-Network (DQN) Architecture for Groundwater Well Testing Decision-Making The diagram illustrates the structure of the Deep Q-Network (DQN) implemented to optimize groundwater well-testing decisions for 561 Irish agents. The model processes the current state of the environment (e.g., demographic, behavioral, and well-related features) through an input layer, followed by four hidden layers with varying numbers of neurons (64, 128, 256, and 512). Using a Q-value estimation, the output layer determines the optimal action, where a = 1 represents testing, and a = 0 represents not testing. Actions are selected to maximize future rewards based on the Bellman equation, with rewards ($r_t$) and updated states ($s_t$+1) provided by the seasonal environment. Feedback loops enable agents to learn and adapt their policies over time to optimize their decision-making strategies.

#### 4) DQN Training

The Mesa library was used to develop ABM in Python, enabling the simulation of agent interactions within a dynamic environment. To optimize agent decision-making, the Deep Q-Network (DQN) training process follows a structured approach, allowing agents to learn optimal strategies efficiently. At the start of each episode, the environment resets to its initial state, and the total reward is set to zero. Actions are then selected using the epsilon-greedy policy, which balances exploitation and exploration. Exploitation involves selecting actions that maximize the expected reward based on the agent's current knowledge. However, relying solely on exploitation limits the agent's ability to explore new strategies, restricting its coverage of the state-action space and potentially leading to suboptimal performance. To counter this, exploration is incorporated by occasionally selecting random actions, allowing the agent to discover unvisited state-action pairs. This enhances the agent's understanding of the environment and prevents it from being confined to locally optimal solutions. The epsilon-greedy strategy governs this balance, where the agent selects a random action with probability $\epsilon$ and chooses the action that maximizes the Q-value with probability **($1-\epsilon$)**. Initially, $\epsilon$ is set to 0.9, encouraging exploration. As training progresses, $\epsilon$ decays exponentially toward zero, shifting the focus from exploration to exploitation. The decay follows an exponential decay factor set to **$10^5$**. After selecting an action, the agent executes it in the environment, receives a reward, and transitions to a new state. The resulting experience, defined as **($s_n$, $a_n$, $r_n$, $s_{n+1}$, done)**, is stored in a replay memory with a capacity of 100,000 experiences. This memory enables diverse training by allowing agents to sample past interactions randomly. To incorporate **agent-to-agent interactions**, multiple agents operate in a shared environment, where one agent's actions influence both its own state transition and that of others. These interactions generate multi-agent experiences stored in the replay memory, capturing both **agent-environment** and **agent-agent** interactions as described in Section 3.4. This approach enriches the diversity of experiences and enhances the robustness of learned decision-making strategies.

Once enough experience is collected, a mini-batch of size 64 is randomly sampled for training. The target Q-values for these experiences are computed using the Bellman update process, which iteratively refines value estimates based on expected cumulative rewards. A loss function based on Mean Squared Error (MSE) was applied, and backpropagation minimized the loss by adjusting network weights through the Adam optimizer with L2 regularization. The model was trained using dropout rates (0.20 to 0.15), learning rates ranging from 0.4 to 0.0001 (0.4, 0.3, 0.2, 0.1, 0.001, 0.0001), and a momentum value of 0.9. A **learning rate scheduler** adjusts the learning rate every **100 steps** to improve training efficiency, allowing the model to fine-tune its weights during later stages. The **target Q-network** is updated every **10,000 steps** by synchronizing weights with the primary Q-network, ensuring training stability and consistent convergence. The model was trained for **over 2000 episodes** to enhance performance and refine the agent's well-testing decision-making capabilities. Once stability was achieved within the DQN framework, the trained baseline models, using the top 20 features, were applied to analyze 14 hypothetical scenarios, enabling the development of a parameterized Agent-Based Model (ABM). This structured approach supports scenario-based investigations and ensures the DQN framework stabilizes training while preventing overfitting by integrating **epsilon-greedy exploration, experience replay, mini-batch learning, and advanced optimization techniques**.

### 3.7. *Scenario weights:*

Weights were assigned to model scenarios based on bivariate statistical associations identified in the survey respondent cohort as well as pooled data from broader national studies. Assigned weights for both response variables (historical well testing and annual testing) are outlined in Table II. The rationale for weight adjustments was predicted on the identified degree of significance (95% v. 99% CI) in addition to evidentiary data and expert opinion. For instance, it was deemed that evidence of a significant relationship between historical well testing and recent occurrence of gastrointestinal illness ($\chi^2$ = 5.025, $p$ = 0.025) found in [12], in consonance with existing national epidemiological data, would strengthen the persuasive power of health-based messaging alone relative to generic risk communication interventions. These 14 hypothetical scenarios were tested on trained ABMs using the top 20 features.

**Table 2:** Weights assigned to investigated intervention scenarios

| Scenario no. | Hypothesis (weight assigned) for historical well testing |
|---|---|
| **Historical well testing (yes/no)** | |
| **1. Incentivised well testing** | Free well testing will increase the probability of well testing (0.4) |
| **2. Free well testing** | Free well testing will increase the probability of well testing (0.9) |
| **3. Household health risk messaging** | Increased messaging about household health risks and the use of substantive case studies will increase the probability of well testing (0.3) |
| **4. Domestic wastewater treatment system messaging** | Increased messaging about DWWTS contamination risks and the use of substantive case studies will increase the probability of well testing (0.2) |
| **5. Implementation of information campaign** | Increased provision and communication of information relating to well testing and contamination risk will increase the probability of well testing (0.4) |
| **6. Adjusting peer influence** | Altering descriptive/injunctive norms and credence assigned to peer advice will increase probability of well testing (0.4) |
| **7. Regulation** | Regulatory changes (e.g., widening of DWWTS inspection/enforcement protocols and introduction of DWWTS tests during property transactions) will increase the probability of well testing (0.7) |
| **8. Free well testing + intensive information campaign** | Free well testing combined with a thorough risk communication campaign will increase the probability of well testing (0.9 + 0.4) |
| **9. Free well testing + regulation** | Free well testing combined with regulation will increase the probability of well testing (0.9 + 0.7) |
| **10. Gender-focused messaging** | Hypothesis: Increasing awareness-raising activities for females and highlighting the risks of extreme weather events for males will increase well testing (0.4) |
| **Annual well testing (yes/no)** | |
| **1. Messaging about rainfall impacts** | Messaging outlining the impacts of heavy rainfall on potential supply contamination raises probability of annual testing (0.2) |
| **2. Index of test result** | Provision of contamination positive test result raises the probability of annual testing (0.2) |
| **3. Messaging about regular maintenance** | Improved messaging outlining the importance of testing as part of regular maintenance raises probability of annual testing (0.2) |
| **4. Implementation of information campaign** | Increased provision and communication of information relating to well testing and contamination risk will increase the probability of annual testing (0.4) |

## 4. PERFORMANCE METRICS:

To evaluate model performance, decision-making performance, and learning efficiency, the following metrics are considered:

- **Epsilon Decay:** Monitors the shift from exploration to exploitation for balanced learning.
- **Learning Efficiency:** Measures how quickly the model adapts to optimal decision-making based on convergence speed, training iterations.
- **Cumulative Reward:** Represents the accumulated reward over time, reflecting overall model predictive decision-making performance. It is computed as represented in Equation (1) [18]:

$$R_t = \sum^{T} \gamma^k rt+k \qquad (1)$$

$$K=0$$

- **Mean Squared Error (MSE):** Assesses Predictive Performance, where a lower MSE indicates better performance as illustrated in Equation (2) [19]:

$$MSE=1/N\sum_{i=1}^{N}(yi-y\hat{}i) \quad (2)$$

## 5. Results and Discussion

### *5.1. Survey completion*

The survey was initiated by 765 private well users, 74.8% (*n* = 572) of whom undertook the online survey and 25.2% (*n* = 193) of whom undertook the physical survey. Survey responses were deemed suitable for analysis, where respondents answered all questions necessary for awareness quantification and subsequent analysis. A total of 560 surveys were retained after the removal of invalid responses, with respondents deriving from all 26 counties in the ROI. A summary of respondent socio-demographics and supply use characteristics are presented in Table III.

### 5.2. Feature Selection and SHAP-Based Interpretation:

The feature sets were incrementally selected from the top 10 to the top 90 using Recursive Feature Elimination (RFE) and analyzed through Explainable AI (XAI) with SHAP. The interpretation demonstrated that the top 20 features yielded the highest predictive performance, underscoring their critical role in enhancing the model's decision-making process.

#### 5.2.1 SHAP-Based Analysis of the Top 20 Most Influential Features Among 90 Attributes

Figures 4 and 5 collectively present a SHAP summary plot and SHAP dot plot, highlighting the top 20 features influencing the model’s predictions, ranked by their importance scores and visualized through their positive contributions. The color gradient from blue (low feature values) to red (high feature values) in Figure 5 visually represents how different feature values influence well-testing decisions. The most significant feature, Information-Seeking Behaviour (0.14), indicates that individuals actively seeking information are more likely to engage in well testing, with higher values (red) in the SHAP dot plot reinforcing its strong contribution. Overall, Well Awareness (0.12) follows, demonstrating that individuals with greater awareness levels tend to test their wells more frequently, with the dot plot confirming that higher awareness values positively impact predictions. The presence of a Treatment System (0.11) and Confidence in the Ability to Maintain a Well (0.11) further emphasize that well infrastructure and self-efficacy contribute to proactive testing, as red values indicate that individuals with treatment systems and higher confidence are more likely to engage in well testing.

Environmental risk perception and external barriers also play a crucial role in shaping well-testing behavior. Total Number of Barriers (0.09) reflects financial, logistical, and accessibility constraints, where higher values correspond to increased testing due to heightened awareness of obstacles. Similarly, EWE Impact Consequences (0.08), Climate Change Impact Concern (0.08), and EWE Impact Likelihood (0.07) highlight the importance of climate and extreme weather-related risks, with red values in the SHAP dot plot indicating that individuals perceiving these risks as significant are more inclined to conduct well testing.

Structural and socioeconomic characteristics further influence decision-making. Income (0.06), Well Age (0.06), and Well Depth (0.05) show that financial stability, older wells, and deeper wells encourage frequent testing due to contamination concerns, which is reflected in the dot plot, where higher values (red) are linked to increased testing behavior. Local Flood History Importance (0.05) reinforces that past flood experiences contribute to greater water safety awareness, and the dot plot confirms that individuals in flood-prone areas exhibit a higher frequency of testing. Overall, EWE Risk Perception (0.04) and EWE Impact Severity (0.04) demonstrate that individuals who recognize environmental risks actively engage in preventive testing, with higher risk perception values positively influencing model predictions.

Demographic and tenure-related factors also contribute significantly to well-testing decisions. Tenure with Well (0.04), Age (0.04), and Residential Tenure (0.04) indicate that well ownership experience, long-term residency, and age foster a greater sense of responsibility for water quality, with the dot plot showing that individuals with longer tenure and stable residence are more likely to engage in well testing. Well Status Awareness (0.03) highlights that individuals who actively monitor their well’s condition take proactive testing measures, as confirmed by the SHAP dot plot’s red values, which indicate that higher awareness leads to increased testing. Education (0.03) and Province (0.02) still contribute positively, reflecting the role of knowledge and regional variations in shaping well-testing behavior, with higher education and region-specific values associated with increased testing likelihood.

These insights from Figures 4 and 5 collectively reinforce how a combination of awareness, infrastructure, environmental perception, and socio-demographic factors enhances the model’s predictive performance. The interpretation of SHAP values across 90 features highlights that these top 20 features significantly improve the model’s predictive performance, ensuring a more robust and well-informed understanding of well-testing behavior.

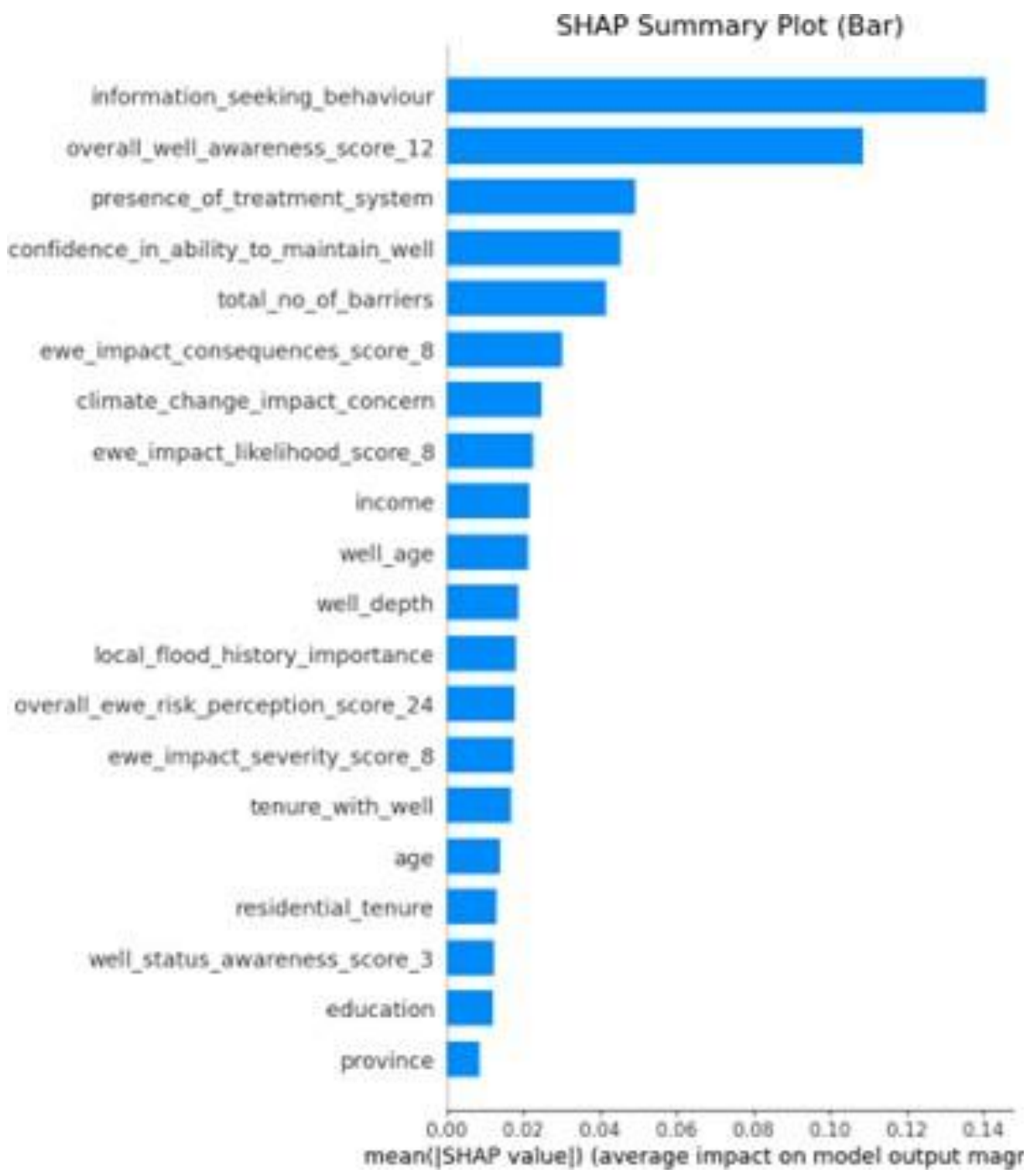

**Figure 4**. SHAP summary plot highlighting the top 20 most influential features and their corresponding impact on Testing (0/1) model predictions.

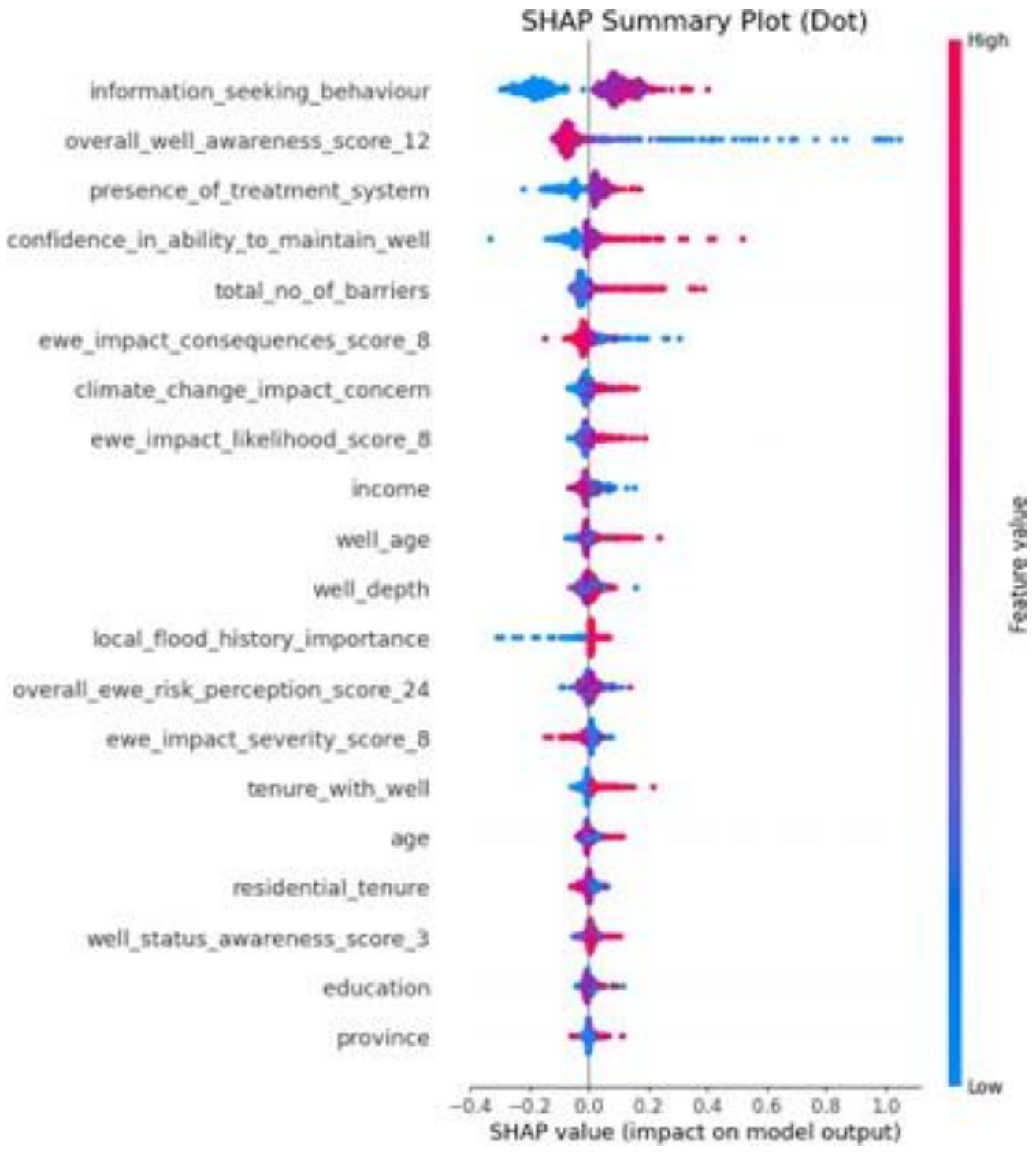

**Figure 5.** SHAP summary dot plot visualizing the top 20 features and their relative importance (impact) on model predictions. Each dot represents the contribution of a feature per agent to the model's output.

### 5.3. DQN ABM Base-Model: Agents Decision-making Performance in Well-Testing (Yes/No)

The Deep Q-Network (DQN) base model (Testing 1/0) was trained on nine feature sets, ranging from 10 to 90 attributes. Using the top 20 features, the model required 2000 episodes to achieve an optimal decision-making performance of 74.10%, with 415 agents starting well testing. This indicates relatively lower learning efficiency as it took more iterations to converge. It achieved this with an MSE of 0.0209, utilizing four hidden layers (64, 128, 256, and 512 neurons), a dropout rate 0.20, and the Adam optimizer as represented in Table 3. In comparison, the top 10 features resulted in a slightly lower decision-making of 73.40%, with a higher MSE of 0.7131 and 410 agents performed well testing.

**Table 3.** DQN BASE-MODEL TESTING (1/0) DECISION-MAKING PERFORMANCE EVALUATION ACROSS TOP 10 AND 20 FEATURES

| Feature Sets (RFE) | Episodes | Hidden Neurons: Layer 1, 2, 3, 4 | Dropout Rate | Learning Rate | Momentum | Optimizer | Decision-making performance (%) | MSE Error | Total Agents performing Testing |
|---|---|---|---|---|---|---|---|---|---|
| Top 10 features | 2000 | 64, 128, 256, 512 | 0.20 | 0.001 | 0.9 | Adam | 73.40 | 0.7131 | 410 |
| Top 20 features | 2000 | - | 0.20 | 0.1 | 0.9 | - | 74.10 | 0.0209 | 415 |

### 5.4. 14 Hypothetical Scenarios Testing:

Table 4 presents the testing results of 14 Irish scenarios, analyzing the impact of various parameters on well-testing outcomes. These scenarios were implemented on a trained baseline model (Testing 1/0) and tested across different frequencies (1, 2, 3, and 4) using the top 20 features. These scenarios are categorized into two sets of testing parameters: Testing Adoption (Y/N) Parameters and Annual Testing (Y/N) Parameters. The Testing Adoption (Y/N) Parameters Set and Annual Testing (Y/N) Parameters Set evaluate different intervention strategies for well-testing, focusing on decision-making performance, mean squared error (MSE), and learning efficiency.

#### 5.4.1 Testing of 14 Hypothetical Scenarios Using (Testing Yes/No) ABM Model 1: Learning Efficiency and Behavioral Analysis

Table 4 shows the Scenario 1 (Incentivized Well Testing), with a probability value of 0.4, involved 421 agents engaged in testing, achieving 75.10% decision-making performance with an MSE of 1.0899. The model exhibited slow learning efficiency, requiring 2000 episodes to converge. Scenario 2 (Free Well Testing), with a probability of 0.9, had 430 agents participating in testing, achieving 76.67% performance with an MSE of 0.2767. This scenario showed faster learning efficiency as agents converged within 1000 episodes, adapting more effectively to well-testing decision-making. Scenario 3 (Household Health Risk Messaging), with a probability of 0.3, included 429 agents conducting tests, yielding 76.45% performance with an MSE of 0.0936. However, the model required 2000 episodes to achieve convergence, indicating slow learning efficiency. Scenario 4 (Domestic Wastewater Treatment System Messaging), with a probability of 0.2, saw 428 agents undergoing testing, achieving 76.23% performance with an MSE of 0.0074, yet requiring 2000 episodes, demonstrating slow adaptation.

Scenario 5 (Implementation of Information Campaigns), with a probability of 0.4, included 426 agents taking part in testing, achieving 75.89% performance with an MSE of 0.0645. However, agents took 2000 episodes to converge, showing slow learning efficiency. Scenario 6 (Adjusting Peer Influence), with a probability of 0.4, involved 423 agents performing well-testing, reaching 75.34% performance with an MSE of 0.6464, yet required 2000 episodes to stabilize decision-making. Scenario 7 (Regulation), with a probability of 0.7, comprised 424 agents participating in tests, achieving 75.53% performance with an MSE of 0.3770, but required 2000 episodes, further reinforcing slow learning efficiency. Scenario 8 (Free Well Testing + Risk Communication Campaign), with combined probability values of 0.9 + 0.4, engaged 435 agents in well-testing and exhibited the highest decision-making performance of 77.23%, with an MSE of 1.1794, while requiring only 1000 episodes, demonstrating faster learning efficiency**.** Scenario 9 (Free Well Testing + Regulation), combining probabilities of 0.9 + 0.7, had 433 agents participating in decision-based testing, attaining 77.11% performance with an MSE of 0.9332, while reaching convergence in 1000 episodes, classifying it as faster in learning efficiency. Scenario 10 (Gender-Focused Messaging), with a probability of 0.4, included 426 agents involved in the testing process, achieving 75.89% performance with an MSE of 0.0645, yet required 2000 episodes, classifying it as another slow-learning model.

The Annual Testing (Y/N) Parameters Set evaluates the effectiveness of long-term interventions in well-testing decision-making. Scenario 11 (Messaging About Rainfall Impacts), with a probability of 0.2, engaged 427 agents actively testing, achieving 76.12% performance with an MSE of 0.2025, but exhibited slow learning efficiency, requiring 2000 episodes. Similarly, Scenario 12 (Index of Test Results), with a probability of 0.2, saw 422 agents participating in assessments, yielding 75.15% performance with an MSE of 0.2704, yet required 2000 episodes to stabilize agent decision-making as illustrated in Table 4. Scenario 13 (Messaging About Regular Maintenance), with a probability of 0.2, included 422 agents involved in test-based decisions, achieving 75.18% performance with an MSE of 0.2401, while also requiring 2000 episodes, confirming slow learning efficiency. Lastly, Scenario 14 (Implementation of Information Campaigns - Annual Testing), with a probability of 0.4, included 428 agents engaged in periodic testing, achieving 76.23% performance with an MSE of 0.3136, yet required 2000 episodes, making it another slow-learning model.

The left plot in Figure 6 illustrates the relationship between training steps, epsilon decay, and average reward score across 1000 episodes for Scenario 8 (Free Well Testing + Risk Communication Campaign). The model, trained on 20 features, achieved an average reward score of 500, indicating faster learning efficiency. The reward score represents the cumulative reward obtained by the agent, reflecting its ability to optimize well-testing decisions over time. A higher reward 500 score confirms the agent's ability to select actions that maximize testing adoption, with steady learning improvement throughout training.

The exploration-exploitation tradeoff is evident in this plot. Initially, the model operates in a high-exploration phase, where epsilon is close to 1, causing the agent to randomly test actions instead of relying on its learned policy. This phase helps the agent evaluate different decisions before epsilon steadily decays (blue line), signaling the shift toward exploitation. By later episodes, the agent relies on optimized decision-making, leading to higher and more stable reward scores. The orange scatter points show a clear upward trend, confirming continuous learning. As epsilon approaches zero, reward scores stabilize, indicating that the agent has successfully optimized its decision process and no longer depends on random exploration.

The right plot in Figure 6 highlights the 435 agents performing well-testing in Scenario 8 over 1000 episodes. Initially, participation is lower, but as training progresses, agent involvement increases, reflecting model stability and improved decision-making confidence. After 900 episodes, participation stabilizes, marking the model's transition to an optimal decision-making phase. The decision-making performance reaches 77.23%, the highest among all scenarios, reinforcing its superior learning efficiency. The consistent rise in efficiency throughout training suggests that integrating free well testing with a risk communication campaign significantly enhances the agent's ability to make faster and more informed decisions, leading to effective well-testing adoption.

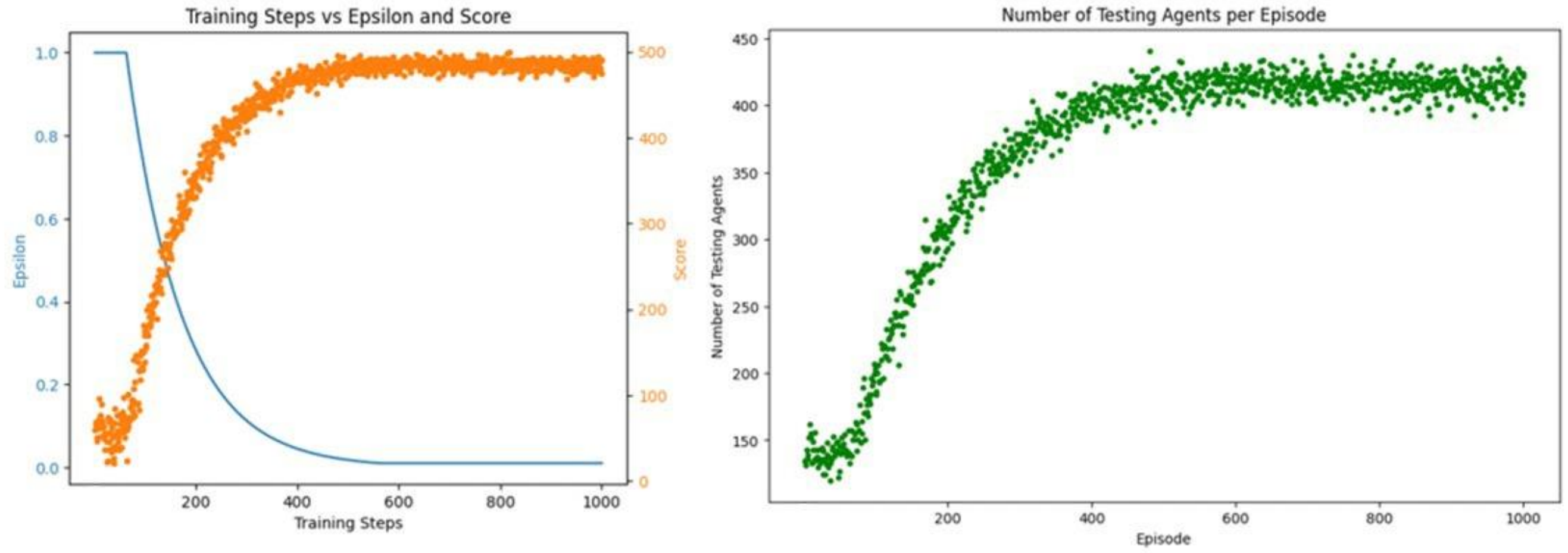

**Figure 6.** Relationship Between Training Steps, Epsilon Decay, and Average Reward Score Across 1000 Episodes for Scenario 8 (Free Well Testing + Risk Communication Campaign). The Left Plot Shows the Epsilon Decay (Blue Line) and Reward Score Trend (Orange Scatter Points), Indicating the Exploration-Exploitation Tradeoff and the Model's Learning Efficiency. The Right Plot Represents the Number of Agents Performing Well-Testing Over Time, stabilizing at 435 Agents, with Decision-Making Performance Reaching 77.23%.

**TABLE 4**. Performance of 14 Hypothetical Scenarios for Well-Testing Behavior, Presenting Total Agents Performing Testing, Decision-Making Performance (%), and Mean Squared Error (MSE). The Results Are Categorized into Testing Adoption and Annual Testing Parameters, with Learning Efficiency Evaluated Based on the Number of Episodes Required for Convergence.

| Scenarios | Parameters | Total Agents Performing Testing | Episodes (Learning Efficiency) | Decision making performance (%) | MSE Error |
|---|---|---|---|---|---|
| **Testing Adoption (Y/N) Parameters:** | | | | | |
| SCENARIO 1 | INCENTIVISED WELL TESTING (Probability: 0.4) | 421 | 2000 | 75.10 | 1.0899 |
| SCENARIO 2 | **FREE WELL TESTING (Probability: 0.9)** | **430** | **1000** | **76.67** | **0.2767** |
| SCENARIO 3 | HOUSEHOLD HEALTH RISK MESSAGING (Probability: 0.3) | 429 | 2000 | 76.45 | 0.0936 |
| SCENARIO 4 | DOMESTIC WASTEWATER TREATMENT SYSTEM MESSAGING (Probability: 0.2) | 428 | 2000 | 76.23 | 0.0074 |
| SCENARIO 5 | IMPLEMENTATION OF INFORMATION CAMPAIGNS (Probability: 0.4) | 426 | 2000 | 75.89 | 0.0645 |
| SCENARIO 6 | ADJUSTING PEER INFLUENCE (Probability: 0.4) | 423 | 2000 | 75.34 | 0.6464 |
| SCENARIO 7 | REGULATION (Probability: 0.7) | 424 | 2000 | 75.53 | 0.3770 |
| SCENARIO 8 | **FREE WELL TESTING + RISK COMMUNICATION CAMPAIGN (Probability: 0.9+0.4)** | **435** | **1000** | **77.23** | **1.1794** |
| SCENARIO 9 | **FREE WELL TESTING + REGULATION (Probability: 0.9 + 0.7)** | **433** | **1000** | **77.11** | **0.9332** |
| SCENARIO 10 | GENDER-FOCUSED MESSAGING (Probability: 0.4) | 426 | 2000 | 75.89 | 0.0645 |
| **Annual Testing (Y/N) Parameters:** | | | | | |
| SCENARIO 1 | MESSAGING ABOUT RAINFALL IMPACTS (Probability: 0.2) | 427 | 2000 | 76.12 | 0.2025 |
| SCENARIO 2 | INDEX OF TEST RESULTS (Probability: 0.2) | 422 | 2000 | 75.15 | 0.2704 |
| SCENARIO 3 | MESSAGING ABOUT Regular MAINTENANCE (Probability: 0.2) | 422 | 2000 | 75.18 | 0.2401 |
| SCENARIO 4 | IMPLEMENTATION OF INFORMATION CAMPAIGNS (Probability: 0.4) | 428 | 2000 | 76.23 | 0.3136 |

### 5.4.2 Evaluation of 14 Hypothetical Scenarios Using an Agent-Based Model 2 (ABM): Testing Frequencies, Learning Efficiency, and Seasonal Variations

**Testing Adoption (Y/N) Parameters** analyze agent participation across 10 scenarios, considering testing frequencies, seasonal variations, and learning efficiencies as illustrated in Table 5.

In **Scenario 1** (Incentivized Well Testing, Probability: 0.4, Slow Learning - 2000 episodes), 421 agents engaged in testing, with 22 conducting tests multiple times per year, 62 opting for annual testing, and 211 performing tests once in a lifetime. This scenario required a prolonged adaptation period as agents took longer to respond to incentives. Seasonally, 226 agents tested in autumn, 107 in spring, 69 in winter, and 19 in summer, indicating a preference for testing in cooler months. Similarly, **Scenario 2** (Free Well Testing, Probability: 0.9, Faster Learning - 1500 episodes) showed higher engagement, with 430 agents participating. Among them, 335 agents tested frequently per year, 65 conducted tests annually, and 17 opted for once-in-a-lifetime testing. The introduction of free well testing resulted in a faster learning pace, as participation increased but still required behavioral reinforcement over time. Here, testing peaked in autumn (235 agents) and spring (106), with lower participation in winter (75) and summer (14).

Health-based messaging in **Scenario 3** (Household Health Risk Messaging, Probability: 0.3, Slow Learning - 2000 episodes) led to 429 individuals participating, but only 11 engaged in frequent testing, 74 tested annually, and 208 opted for rare testing. The slow learning curve in this scenario suggests that awareness alone was not an immediate motivator for behavioral change. Seasonally, testing engagement was higher in autumn (204 agents) and spring (109), while winter (71) and summer (45) showed less participation. A similar pattern was observed in **Scenario 4** (Domestic Wastewater Treatment System Messaging, Probability: 0.2, Slow Learning - 2000 episodes), which had 428 agents participating, with 19 testing frequently, 53 testing annually, and 195 testing once in a lifetime. The learning rate remained slow, as messaging efforts required continuous exposure before behavioral shifts occurred. Seasonal participation followed a similar trend, with higher engagement in autumn (223 agents) and spring (108), while winter (62) and summer (35) saw lower testing activity.
In Scenario 5 (Implementation of Information Campaigns, Probability: 0.4, Slow Learning - 2000 episodes), 426 agents engaged in testing, where 14 tested frequently, 55 tested annually, and 209 opted for once-in-a-lifetime testing. The impact of such campaigns was gradual, as agents required repeated exposure to change their habits. The seasonal variation showed higher testing in autumn (216 agents) and spring (104), compared to winter (69) and summer (37). Similarly, **Scenario 6** (Adjusting Peer Influence, Probability: 0.4, Slow Learning - 2000 episodes) involved 423 agents, including 23 who tested frequently, 66 who opted for annual testing, and 216 who conducted tests once in a lifetime. The influence of social networks contributed to a slow but steady learning process, where adoption depended on broader community acceptance. Here, 228 agents tested in autumn, 114 in spring, 68 in winter, and 13 in summer.

Regulatory enforcement in **Scenario 7** (Regulation, Probability: 0.7, Slow Learning - 2000 episodes) recorded 424 agents performing tests, with 25 testing frequently, 70 opting for annual testing, and 203 testing once in a lifetime. The implementation of regulations required a long-term enforcement period before widespread compliance was observed. Seasonal trends showed higher engagement in autumn (230 agents) and spring (117), with lower participation in winter (64) and summer (13). However, a more effective approach was seen in **Scenario 8** (Free Well Testing + Risk Communication Campaign, Probability: 0.9 + 0.4, Fast Learning - 1000 episodes), which exhibited rapid learning efficiency. A total of 435 agents engaged in testing, where 346 performed tests multiple times per year, 73 opted for annual testing, and 13 tested once in a lifetime, as displayed in Figure 7. The combination of free testing and active communication accelerated the learning rate, allowing for faster adoption within 1000 episodes. Seasonally, testing peaked in autumn (242 agents) and spring (107), while winter (77) and summer (9) showed lower participation, as indicated in Figure 8. The testing participation is illustrated in Figure 9 and Figure 10, where Frequency 1 shows an increase in the number of testing agents, stabilizing around 346 agents who perform well testing frequently annually, while Frequency 2 shows an increase, peaking at approximately 73 agents who opt for annual well testing.

In **Scenario 9** (Free Well Testing + Regulation, Probability: 0.9 + 0.7, Quicker Learning - 1500 episodes), 433 agents participated, with 340 testing frequently, 69 opting for annual testing, and 10 testing once in a lifetime. Introducing regulation alongside free testing led to a quicker learning speed, as agents adapted more quickly than in regulation-only settings. Seasonal data showed higher participation in autumn (234 agents) and spring (105), while winter (79) and summer (15) exhibited lower engagement. Lastly, **Scenario 10** (Gender-Focused Messaging, Probability: 0.4, Slow Learning - 2000 episodes) saw 426 agents engaging in testing, where 19 tested frequently, 59 tested annually, and 226 tested once in a lifetime. The effectiveness of gender-specific campaigns was not immediate, requiring long-term exposure before observable changes in behavior occurred. Seasonal participation was highest in autumn (213 agents) and spring (106), with lower engagement in winter (71) and summer (36).

The **Annual Testing (Y/N) Parameters** examine how different strategies influence well-testing participation over time, considering testing frequency, agent engagement, and seasonal trends, as shown in Table 5. In **Scenario 11** (Messaging About Rainfall Impacts, Probability: 0.2, Slow Learning - 2000 episodes), 427 agents participated, where 22 conducted tests multiple times per year, 85 opted for annual testing, and 210 performed testing once in a lifetime. The highest engagement was seen in autumn (213 agents) and spring (106), while winter (71) and summer (37) showed lower participation, indicating that rainfall-related messaging had a moderate influence on decision-making.
In **Scenario 12** (Index of Test Results, Probability: 0.2, Slow Learning - 2000 episodes), 422 agents took part, among whom 20 engaged in frequent testing, 69 conducted annual tests, and 215 tested once in a lifetime. The seasonal trend followed a similar pattern, with higher participation in autumn (211 agents) and spring (105), while winter (70) and summer (36) showed reduced engagement. The availability of an index system encouraged some level of consistency in testing but did not significantly accelerate the adoption rate.

For **Scenario 13** (Messaging About Regular Maintenance, Probability: 0.2, Slow Learning - 2000 episodes), 422 agents engaged in testing, where 10 performed tests frequently, 50 opted for annual testing, and 183 tested only once in a lifetime. The seasonal breakdown showed 211 agents testing in autumn, 105 in spring, 70 in winter, and 36 in summer, indicating that maintenance reminders alone had a limited immediate impact on increasing routine testing behaviors.
In **Scenario 14** (Implementation of Information Campaigns, Probability: 0.4, Slow Learning - 2000 episodes), 428 agents participated, with 11 conducting frequent tests, 48 opting for annual testing, and 212 choosing rare testing. Seasonal participation was highest in autumn (214 agents) and spring (107), while winter (71) and summer (36) had lower engagement, reinforcing that while informational efforts can increase awareness, they take time to drive consistent behavioral change.

**TABLE 5.** Testing Frequency Across 14 Hypothetical Scenarios Based on Seasonal Variations, Learning Efficiency, and Policy Interventions

| **Scenarios** | **Parameters** | **Episodes (Learning Efficiency)** | **Total Agents Performing Testing** | **2-3 Per Year** (a_f=1) | **Once Annually** (a_f=2) | **Once Every Few Years** (a_f=3) | **Once in a Lifetime** (a_f=4) | **Autumn** | **Spring** | **Winter** | **Summer** |
|---|---|---|---|---|---|---|---|---|---|---|---|
| **Testing Adoption (Y/N) Parameters:** | | | | | | | | | | | |
| SCENARIO1 | INCENTIVISED WELL TESTING (Probability: 0.4) | 2000 | 421 | 22 | 62 | 211 | 126 | 226 | 107 | 69 | 19 |
| SCENARIO2 | **FREE WELL TESTING (Probability: 0.9)** | **1000** | **430** | **335** | **65** | **17** | **13** | **235** | **106** | **75** | **14** |
| SCENARIO3 | HOUSEHOLD HEALTH RISK MESSAGING (Probability: 0.3) | 2000 | 429 | 11 | 74 | 208 | 136 | 204 | 109 | 71 | 45 |
| SCENARIO4 | DOMESTIC WASTEWATER TREATMENT SYSTEM MESSAGING (Probability: 0.2) | 2000 | 428 | 19 | 53 | 195 | 161 | 223 | 108 | 62 | 35 |
| SCENARIO5 | IMPLEMENTATION OF INFORMATION CAMPAIGNS (Probability: 0.4) | 2000 | 426 | 14 | 55 | 209 | 148 | 216 | 104 | 69 | 37 |
| SCENARIO6 | ADJUSTING PEER INFLUENCE (Probability: 0.4) | 2000 | 423 | 23 | 66 | 216 | 118 | 228 | 114 | 68 | 13 |
| SCENARIO7 | REGULATION (Probability: 0.7) | 2000 | 424 | 25 | 70 | 203 | 126 | 230 | 117 | 64 | 13 |
| SCENARIO8 | **FREE WELL TESTING + RISK COMMUNICATION CAMPAIGN (Probability: 0.9+0.4)** | **1000** | **435** | **346** | **73** | **13** | **3** | **242** | **107** | **77** | **9** |
| SCENARIO9 | **FREE WELL TESTING + REGULATION (Probability: 0.9 + 0.7)** | **1000** | **433** | **340** | **69** | **10** | **14** | **234** | **105** | **79** | **15** |
| SCENARIO 10 | GENDER-FOCUSED MESSAGING (Probability: 0.4) | 2000 | 426 | 19 | 59 | 226 | 122 | 213 | 106 | 71 | 36 |
| **Annual Testing (Y/N) Parameters:** | | | | | | | | | | | |
| SCENARIO1 | MESSAGING ABOUT RAINFALL IMPACTS (Probability: 0.2) | 2000 | 427 | 22 | 85 | 210 | 110 | 213 | 106 | 71 | 37 |
| SCENARIO2 | INDEX OF TEST RESULTS (Probability: 0.2) | 2000 | 422 | 20 | 69 | 215 | 118 | 211 | 105 | 70 | 36 |
| SCENARIO3 | MESSAGING ABOUT Regular MAINTENANCE (Probability: 0.2) | 2000 | 422 | 10 | 50 | 183 | 179 | 211 | 105 | 70 | 36 |
| SCENARIO4 | IMPLEMENTATION OF INFORMATION CAMPAIGNS (Probability: 0.4) | 2000 | 428 | 11 | 48 | 212 | 157 | 214 | 107 | 71 | 36 |

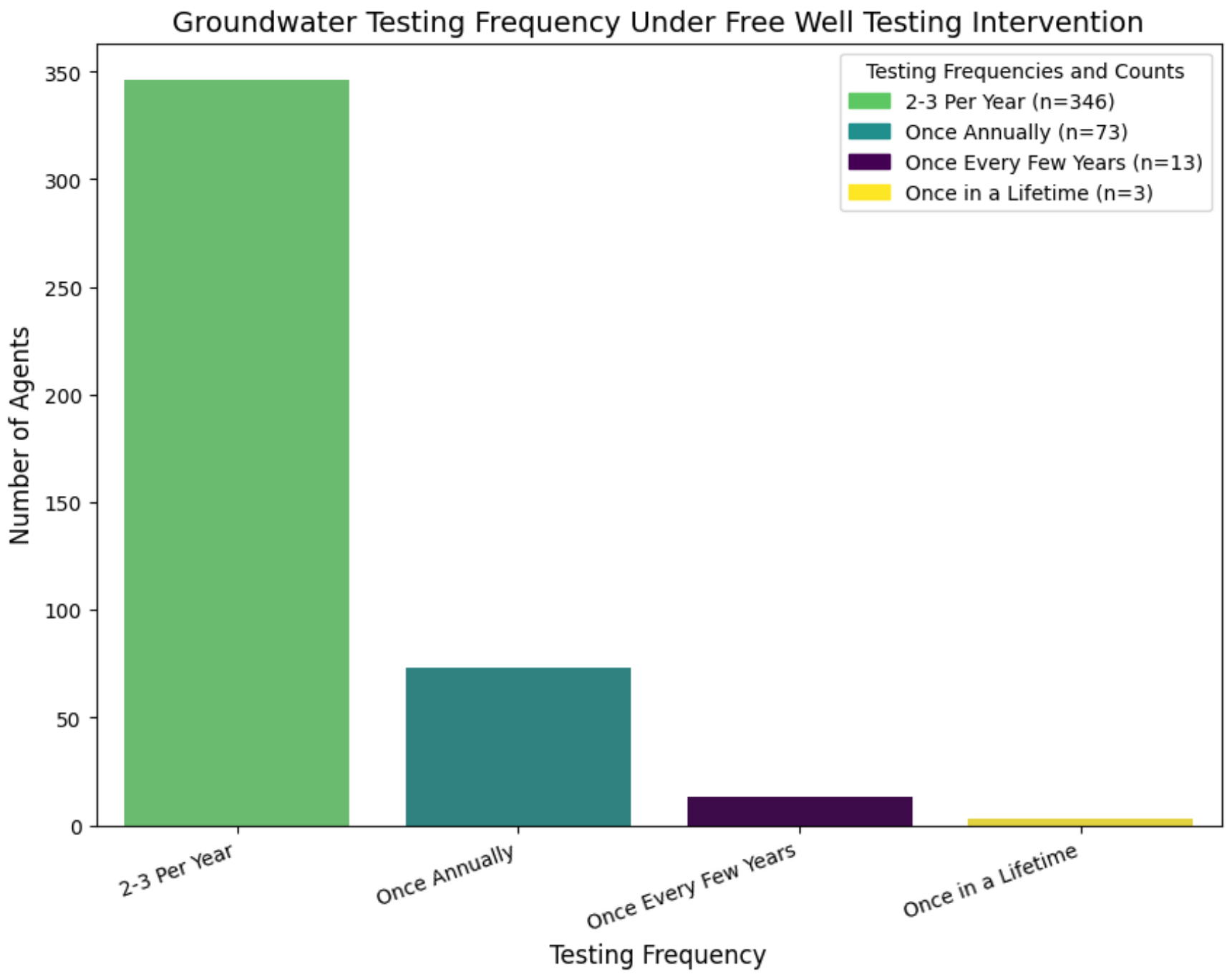

**Figure 7.** Distribution of Agents Performing Groundwater Testing Under the Free Well Testing Scenario. The bar chart illustrates the frequency of groundwater testing among agents, showing that the majority (346 agents) conduct tests 2-3 times per year, while fewer agents opt for annual testing (73), testing once every few years (13), or only a single test in their lifetime (3).

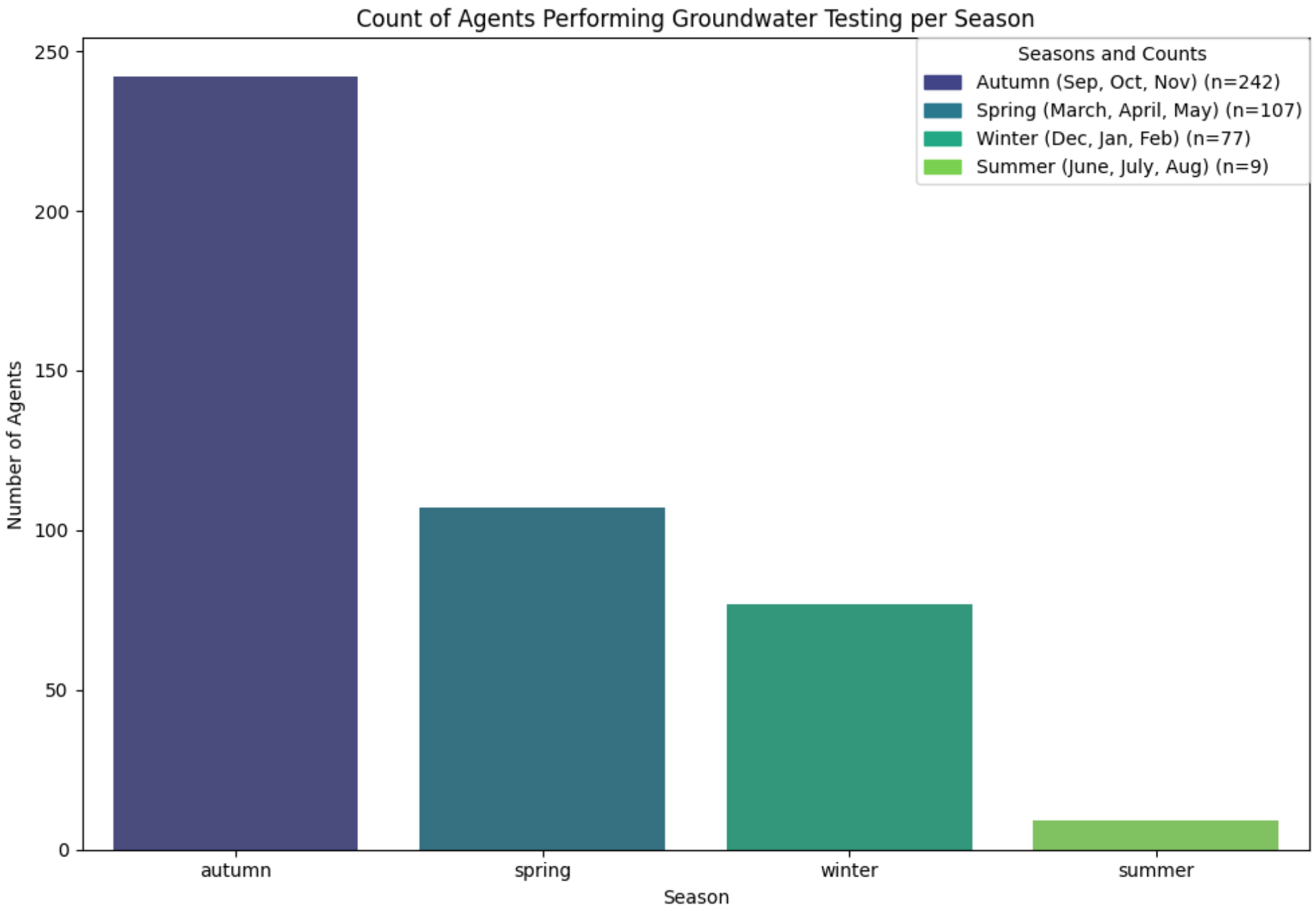

**Figure 8.** Seasonal variation in groundwater testing, with the highest participation in **autumn (242 agents)** and the lowest in **summer (9 agents)**, indicating seasonal influence on testing behavior

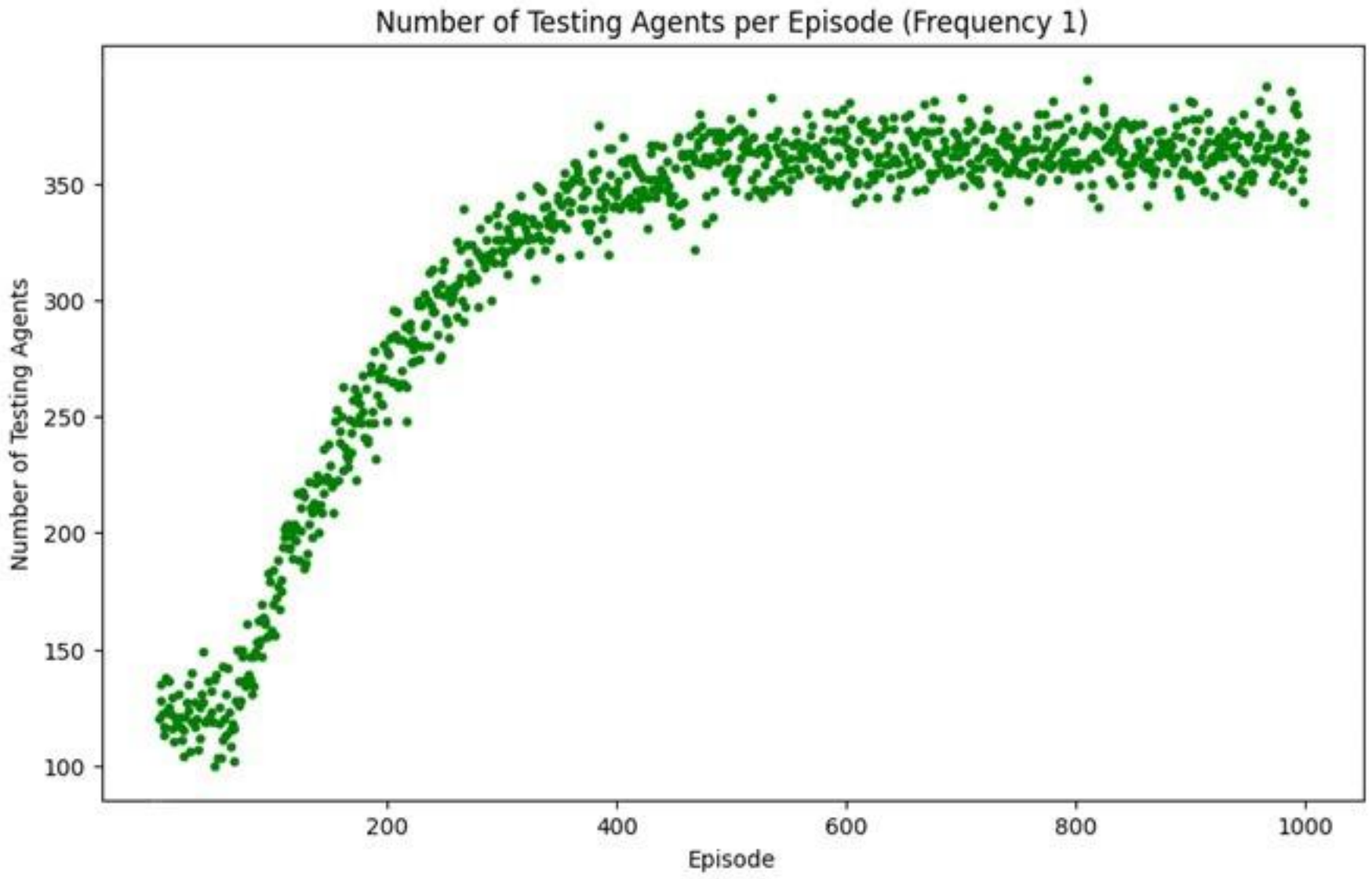

**Figure 9.** The number of agents adopting frequent well testing (multiple times per year) under Scenario 8 (Free Well Testing + Risk Communication Campaign) is over 1000 episodes. The graph shows an increase in participating agents and stabilizes around 346, reflecting accelerated adoption due to free testing and active risk communication.

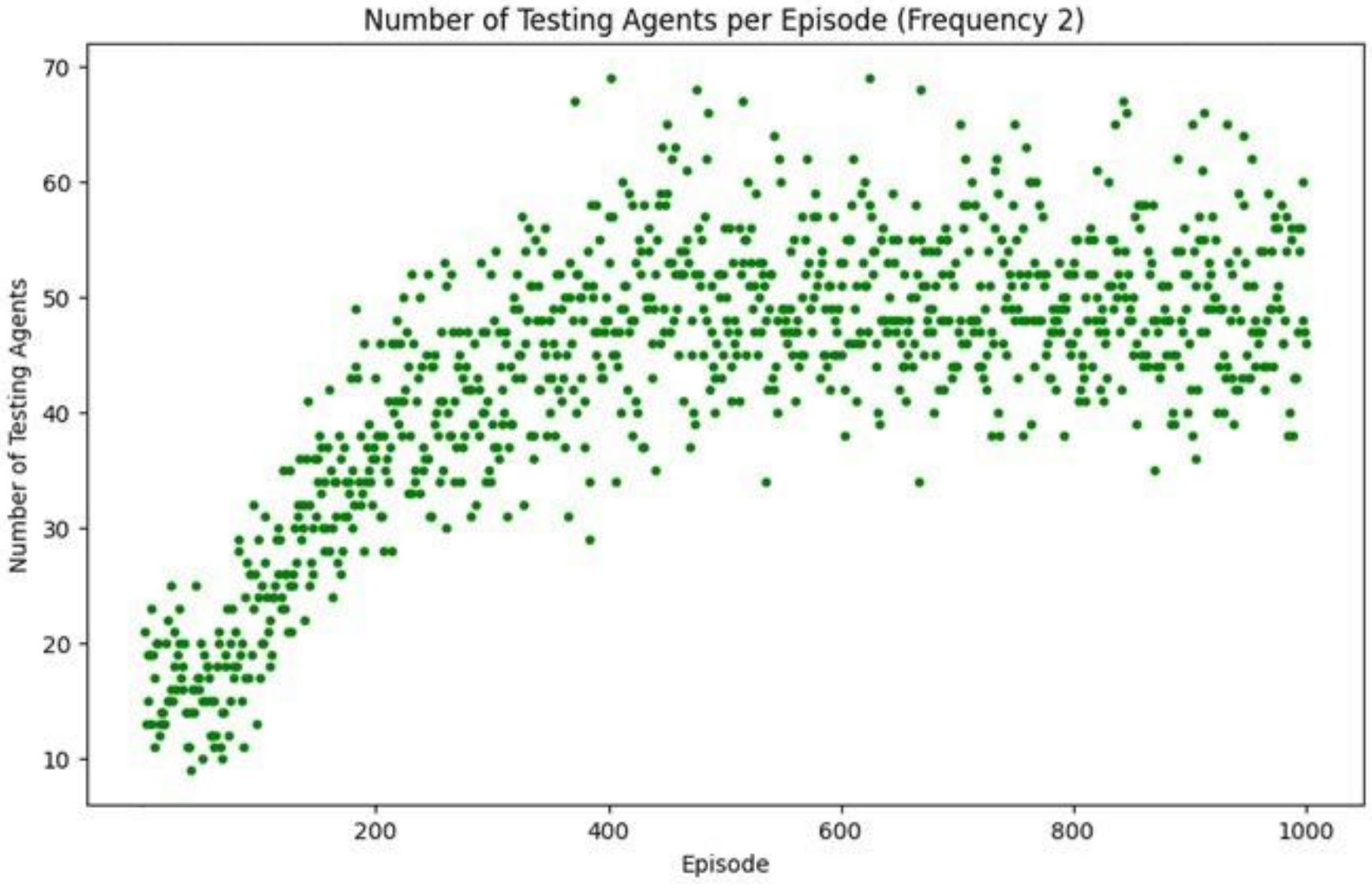

**Figure 10.** The incremental increase in annual well-testing adoption occurs when tests are offered at no cost and risks are clearly explained, ultimately leading 73 agents to perform well-testing annually.

## 6. CONCLUSION

This study presents a novel Agent-Based Modeling (ABM) framework integrated with Deep Q-network (DQN) reinforcement learning and Explainable Artificial Intelligence (XAI) to analyze and optimize private well-testing behaviors in rural Ireland. By simulating 14 policy-driven scenarios, the model provides quantifiable insights into behavioral responses, addressing a critical gap in water safety policy and public health research. Among the simulated interventions, "Free Well Testing + Communication Campaign" emerged as the most effective, leading to 435 out of 561 agents participating in well testing and demonstrating faster learning efficiency within 1000 episodes—a substantial increase from the baseline participation rate of approximately 5%. Similarly, "Free Well Testing + Regulation" proved highly effective, with 433 out of 561 agents initiating well testing and adapting efficiently to the intervention. Additionally, free testing alone significantly increased testing frequency, exceeding 75% participation, with many residents opting for multiple tests annually, reinforcing the effectiveness of financial incentives in encouraging proactive water quality monitoring. The free well-testing scenarios consistently exhibited accelerated learning efficiency, converging within 1000 episodes, highlighting the role of incentive-driven interventions in shaping behavioral change. In contrast, other scenarios required approximately 2000 episodes, taking twice as long for decision-making, indicating slower adaptation rates and prolonged convergence periods. To enhance model interpretability, Recursive Feature Elimination (RFE) with 10-fold cross-validation identified key behavioral predictors, while SHAP (Shapley Additive Explanations), a core XAI technique, provided granular insights into the relative importance of the top 20 features, such as risk perception, water contamination history, testing cost, social norms, government incentives, income level, trust in authorities, past testing behavior, health concerns, media and peer influence, enforcement likelihood, knowledge of contamination risks, and household characteristics. Integrating XAI-driven SHAP values ensures model transparency, improving interpretability and facilitating evidence-based policy formulation.